\documentclass[a4paper, twocolumn]{revtex4} 
\usepackage{color}
\usepackage{graphics}
\usepackage{epsfig}
\begin{document}
	
	\title{Structure and magnetic studies of geometrically frustrated disordered pyrochlores A$_{2}$Zr$_{2}$O$_{7}$: (A = Eu, Gd, Er)}
	
	\author{Sheetal and C. S. Yadav*}
	\affiliation{School of Basic Sciences, Indian Institute of Technology Mandi, Mandi-175075 (H.P.), India}
	
	\begin{abstract}
				
		The spin ice system Dy$_{2}$Ti$_{2}$O$_{7}$ exhibits strong frequency-dependent spin-freezing at $\sim$ 16 K temperature. Although it has been a matter of discussion for years, the origin of this unusual spin freezing is still unknown. The replacement of Ti with isovalent Zr leads to the dynamic magnetic ground state at low temperatures in Dy$_{2}$Zr$_{2}$O$_{7}$ and prevents the formation of high-temperature spin freezing. Interestingly the high-temperature spin freezing re-emerges in the presence of the magnetic field. In this direction, we have studied a series of disordered pyrochlore oxides A$_{2}$Zr$_{2}$O$_{7}$ (A = Eu, Gd, Er) and compared their crystal structure, magnetic, and heat capacity behavior with that of Dy$_{2}$Zr$_{2}$O$_{7}$ and Ho$_{2}$Zr$_{2}$O$_{7}$ systems. Our study shows that depending on the disordered parameter, the spin-freezing behavior can be retained by slowing down the spin dynamic with a suitable choice of the magnetic field.	We observe that unlike titanates, modification at the rare earth site does not make considerable change in the magnetic ground state of these zirconates compounds.  
		
	\end{abstract}
	
	\maketitle
	
	\section{Introduction} 
	
The observation of spin ice state is one of the most notable aspects of the pyrochlore class of compounds. Although the initial excitement in these systems was in the form of magnetic analog of water ice, it was later expanded to include a variety of unusual behaviors like dynamic spin ice \cite{zhou2008dynamic}, ordered spin ice \cite{mirebeau2005ordered}, quantum spin ice \cite{sibille2018experimental} and Kagome ice \cite{fennell2007pinch, tabata2006kagome}. In spin ice materials, the local spin configuration of two-in-two-out maps onto the Bernal-Fowler rule for proton-oxygen bond length in water ice and the associated remnant Pauling residual entropy as T $\rightarrow$ 0 K is the primary signature for the classical spin-ice state. The origin of the spin ice state is explained based on the dipolar spin ice model and has been successfully reproduced the experimental observation of heat capacity and basic features of neutron diffraction studies \cite{isakov2005spin,den2000dipolar,melko2004monte}.\\  
	
The pyrochlore titanates (Ho/Dy)$_{2}$Ti$_{2}$O$_{7}$, has been extensively studied over the past few years to understand the mechanism of low-temperature spin-ice behavior \cite{ramirez1999zero,snyder2001spin,bramwell2001spin}. The magnetic properties of these A$_{2}$B$_{2}$O$_{7}$ pyrochlore systems (where A is the rare-earth ion and B is the transition-metal ion) depend on the choice of A and B atoms. Recently, a disordered pyrochlore zirconate Dy$_{2}$Zr$_{2}$O$_{7}$ have attracted considerable attention because of the absence of spin-ice state on replacing the Ti in Dy$_{2}$Ti$_{2}$O$_{7}$ with the Zr ion due to the subtle interplay between the chemical disorder and spin dynamics \cite{ramon2019absence}. The modification in the structure leads to a very dynamic ground state with an entropy of Rln2 instead of a spin ice state. The suppression of spin ice state shows an intimate connection between the structure and electronic properties in these systems. The crystal field effect of symmetry D$_{3d}$ (oxygen environment around A site (rare-earth)) strongly affects the state of magnetic ions and the magnetic structure they create. Based on the structural data, it has been argued that the modified symmetry of the distorted cube AO$_{8}$ plays an important role in determining the unusual low-temperature magnetic behavior of A$_{2}$Zr$_{2}$O$_{7}$ systems. Further, the replacement of Ti site in Ho$_{2}$Zr$_{2}$O$_{7}$ leads to the absence of spin ice state due to chemical disorder and emergence of field-induced spin freezing at high temperature \cite{sheetal2021field}. The application of field and partial substitution at the magnetic site also give rise to some interesting observations. Dy$_{2}$Zr$_{2}$O$_{7}$ is reported to exhibits the field-induced high-temperature spin freezing, which further leads to the spin ice ground state by possessing the Pauling residual entropy at \textit{H} = 5 kOe, which is a primary signature for spin ice state \cite{sheetal2020emergence}. Further, the nonmagnetic La$^{3+}$ substitution at Dy$^{3+}$ site stabilized pyrochlore structure along with the high-temperature spin freezing without the external magnetic field \cite{sheetal2021evolution}. In addition, the Gd$_{2}$(Ti/Sn)$_{2}$O$_{7}$ shows the magnetically ordered state even with the persistent spin dynamics \cite{chapuis2009probing,brammall2011magnetic,bertin2002effective}. Er$_{2}$Ti$_{2}$O$_{7}$ is an XY antiferromagnet on a pyrochlore lattice and is a candidate material for observing the order by disorder phenomenon leading to an ordered state \cite{de2012magnetic,champion2003er}. Eu$_{2}$Ti$_{2}$O$_{7}$ exhibits unusual high-temperature spin freezing and shows the temperature-independent dc magnetic susceptibility at low temperature due to the mixing of the nonmagnetic ground state of Eu$^{3+}$ with the excited states \cite{pal2018high}. Therefore it is important to look for the signature of high temperature spin freezing in other iso-structural compounds to understand the role of structural deformation and magnetic interactions on the spin freezing state in the spin-ice materials. \\

Here we study the crystal structural, magnetism and the heat capacity of a series of compounds Er$_{2}$Zr$_{2}$O$_{7}$, Gd$_{2}$Zr$_{2}$O$_{7}$, and Eu$_{2}$Zr$_{2}$O$_{7}$. It is to note that Zr ion is chosen over Ti site, which is also a tetravalent nonmagnetic ion, so that not only the structural symmetry but the magnetic ground state also is expected to affected. Unlike the titanates, we found that the effect of changing A site does not make considerable change in the magnetic behavior of zirconates. We measured the ac susceptibility in the presence of a magnetic field, and found the field induced spin freezing for higher field values. In addition, we observed a temperature independent plateau region in the range \textit{T} = 20 - 90 K in the dc susceptibility in case of Eu$_{2}$Zr$_{2}$O$_{7}$, which persist even at high fields due to the Van Vleck contribution. These studies indicate that the large spin dynamics due to the induced structural disorder can be controlled with the external magnetic field to stabilize the new magnetic state.	
	
\section{Experimental Details}
The polycrystalline samples were prepared by standard solid-state chemical route using the high purity constituent oxides (A$_{2}$O$_{3}$ $>$ 99.99$\%$ and ZrO$_{2}$ $>$ 99.9$\%$) in stoichiometric amounts. The rare-earth oxides Gd$_{2}$O$_{3}$, Er$_{2}$O$_{3}$ and Eu$_{2}$O$_{3}$ were pre-heated for several hours at 200$^{o}$ C to avoid any moisture due to the hygroscopic nature of these oxides. After the initial step, several heat treatments with intermediate grindings were given at 1350$^{o}$ C and the final heat treatment for 50 hours was given to pellets at the same temperature. The crystal structure and phase purity of the compounds were confirmed by performing the Rietveld refinement of the x-ray diffraction (XRD) pattern using Fullprof Suit software. Magnetization measurements were performed as a function of temperature (\textit{T}) and magnetic field (\textit{H}) using Quantum Design built Magnetic Property Measurement System (MPMS) in the temperature range 1.8 - 300 K. 
	
	\begin{figure}[ht]
		\begin{center}
			\includegraphics[width=7cm,height=10cm]{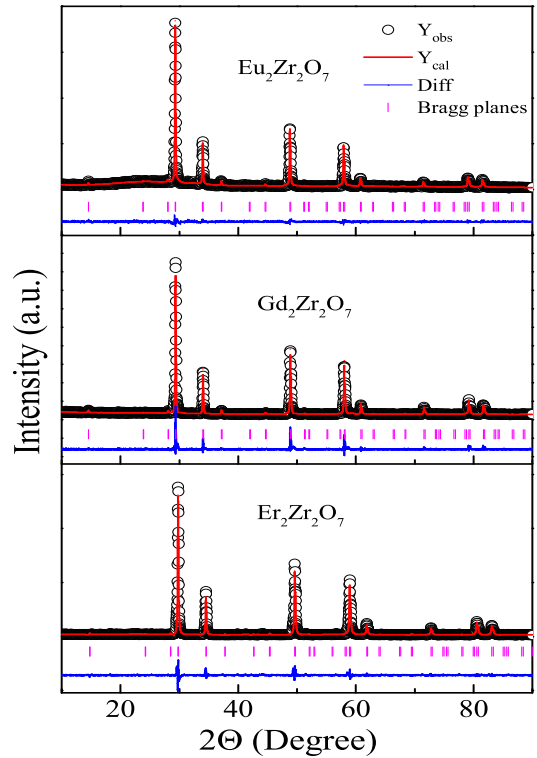}
			\vspace{-5pt}
			\caption{(a) Room temperature Rietveld refined x-ray diffraction pattern of A$_{2}$Zr$_{2}$O$_{7}$ (A = Eu, Gd, Er) with Fd$\bar{3}$m space group.}
		\end{center}
	\end{figure} 
	
	\section{Result and Discussion}
	
	\subsection[A]{Structural studies}
	
	\begin{figure}[ht]
		\begin{center}S
			\includegraphics[width=8.5cm,height=9cm]{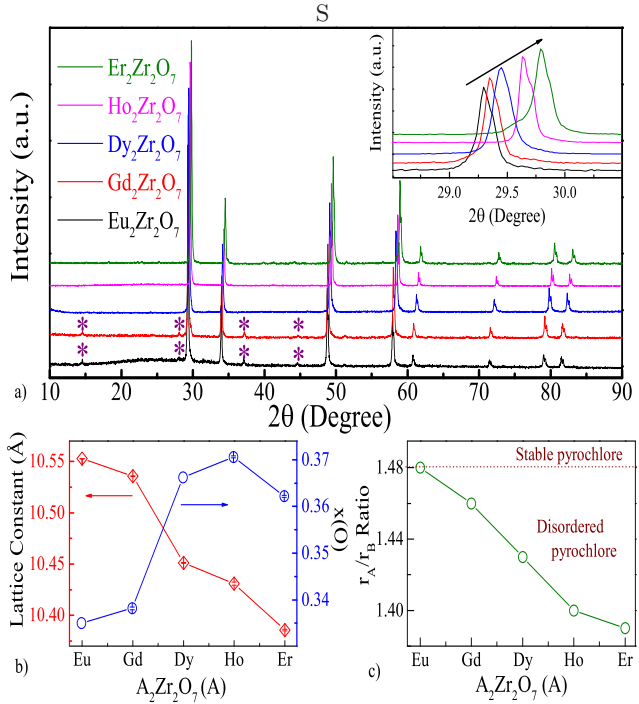}
			\caption{(a) Room temperature Rietveld refined x-ray diffraction pattern of A$_{2}$Zr$_{2}$O$_{7}$ (A = Eu, Gd, Dy, Ho, Er). The xrd data for Ho and Dy zirconate is taken from reference 13 $\&$ 14. Inset: Shows the shift in peak position. A clear shift in peak positions towards higher 2$\theta$ values indicates the decrease in lattice constant. (b) and (c) Shows the variation in lattice constant (left), x(O) parameter (right) and the r$_{A}$/r$_{B}$ ratio for all the series compounds.}
		\end{center}
	\end{figure}
	
	\begin{figure}[ht]
		\begin{center}
			\includegraphics[width=8.5cm,height=3.25cm]{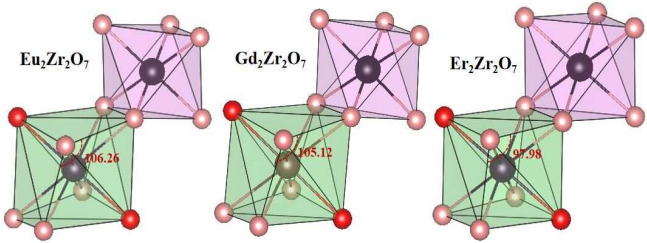}
			\vspace{-5pt}
			\caption{Coordination of A$^{3+}$ (Eu, Gd, Er) and B$^{4+}$ (Zr) ions in A$_{2}$B$_{2}$O$_{7}$. In the structure unit O$^{2-}$ ions surrounds A$^{3+}$ (blue) rare-earth ion and B$^{4+}$ (black) transition-metal ion form the distorted cube and octahedra respectively. The two types of oxygen are shown in red (8d) and light red (48f) color. It shows significanst change in the symmetry of cube about A site, form a comparatively less distorted cube for Eu$_{2}$Zr$_{2}$O$_{7}$ where the O-O-O bond angle between 48f oxygen ion is close to 90$^{o}$.}
		\end{center}
		\vspace{-15pt}
	\end{figure} 
	
	Rietveld refinement of XRD pattern characterized the samples in a single phase without any impurity (see Fig. 1). Modeling of the XRD data with the cubic space group, Fd$\bar{3}$m, consistent with the disordered pyrochlore structure, resulted in excellent fit compared to Fm$\bar{3}$m (fluorite) space group. The crystallographic parameters are summarized in Table 1. We have plotted the XRD pattern of disordered pyrochlore oxides Dy$_{2}$Zr$_{2}$O$_{7}$ and Ho$_{2}$Zr$_{2}$O$_{7}$ also for comparison. X-ray and Raman's studies confirm the formation of disordered pyrochlore phase in these compounds \cite{sheetal2020emergence}. The stability of the pyrochlore structure is determined by two-parameters: radius ratio (r$_{A}$/r$_{B}$) of cations A$^{3+}$ and B$^{4+}$ and the positional parameter x(O) of 48f site of the oxygen atom. For stable pyrochlore the r$_{A}$/r$_{B}$ should lie in the range 1.48 - 1.78 and x(O) $\sim$ 0.3323 \cite{kuznetsov2020magnetization}. Any deviation from these values gives rise to disorder in the pyrochlore lattice or the formation of a new lattice structure fluorite/perovskite. The XRD pattern of A$_{2}$Zr$_{2}$O$_{7}$ (A = Er, Dy, Ho) consists of only the main peaks of pyrochlore structure and the remaining superstructure peaks expected at 2$\theta$ = 14$^{o}$, 27$^{o}$, 36$^{o}$, 42$^{o}$, etc. are missing \cite{mandal2006preparation}. On the other hand, the XRD pattern of A$_{2}$Zr$_{2}$O$_{7}$ (A = Eu, Gd) slightly varies from others with the presence of few superstructure peaks at the lower 2$\theta$ values (marked by a star in Fig. 2), indicating the doubling of the cubic unit cell which is a characteristic of pyrochlore structure. This behavior is consistent with the r$_{A}$/r$_{B}$ ratio of these compounds (see Fig. 2c). In the case of zirconates, the Gd$_{2}$Zr$_{2}$O$_{7}$ is considered at a phase boundary between the pyrochlore and fluorite phase and it can be formed in any of the structural phases depending on the synthesis method \cite{mccauley1973luminescence}. Figure 2b shows the variation in lattice parameter (left) and positional parameter x(O) (right) for all the compounds. The x(O) parameter for Eu and Gd based zirconate lies close to the stable pyrochlore regime (x(O) $\sim$ 0.33) and for others, it lies in the range of disordered pyrochlore (x(O) $\sim$ 0.375). Thus Eu$_{2}$Zr$_{2}$O$_{7}$ and Gd$_{2}$Zr$_{2}$O$_{7}$ adopt a partially ordered pyrochlore lattice compared to the disordered pyrochlores like Er$_{2}$Zr$_{2}$O$_{7}$, Dy$_{2}$Zr$_{2}$O$_{7}$ and Ho$_{2}$Zr$_{2}$O$_{7}$.
	
	\begin{table}[htbp]
		\caption{Crystallographic parameters of A$_{2}$Zr$_{2}$O$_{7}$; A = Eu, Gd, Dy, Ho, Er}
		\begin{tabular}{ccccccccccccc}
			\hline
			A && r$_{A}$/r$_{B}$ && $\chi^{2}$ && a ($\AA$) && x(O) && Ref\\
			\hline
			Eu && 1.48 && 1.23 && 10.5527(3) && 0.3350(7) && This work\\
			Gd && 1.46 && 2.17 && 10.5358(3) && 0.3382(9) && This work\\
			Dy && 1.43 && 1.46 && 10.4511(3) && 0.3667(23) && Ref\cite{sheetal2020emergence}\\
			Ho && 1.40 && 1.86 && 10.4310(2) && 0.3726(17) && Ref\cite{sheetal2021field}\\
			Er && 1.39 && 2.10 && 10.3856(3) && 0.3668(11) && This work\\    	    
			\hline
		\end{tabular}
	\end{table}
	
	\subsection[A]{Magnetic susceptibility}
	
	\begin{figure}[ht]
		\begin{center}
			\includegraphics[width=8cm,height=10cm]{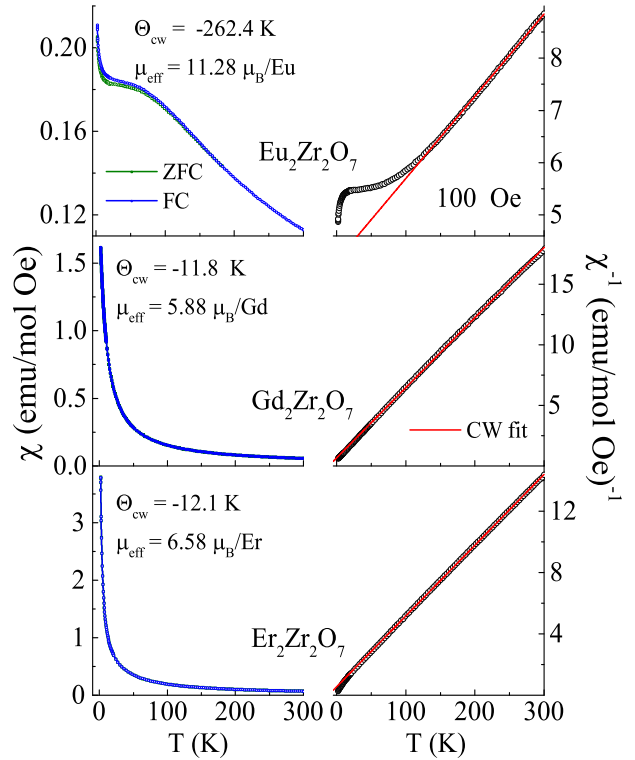}
			\vspace{-5pt}
			\caption{(a) Temperature dependence of magnetic susceptibility (left) and inverse magnetic susceptibility (right) along with the Curie-Weiss fit using $\chi$ = \textit{C}/(\textit{T} - $\theta_{CW}$).}
		\end{center}
		\vspace{-25pt}
	\end{figure} 
	
	DC magnetic susceptibility results for all the three samples in the form of $\chi$ versus \textit{T} and $\chi^{-1}$ versus \textit{T} (1.8 - 300 K in \textit{H} = 100 Oe) are shown in Fig. 4. We did not observe any separation between zero-field cooled (ZFC) and field-cooled (FC) curves in Gd$_{2}$Zr$_{2}$O$_{7}$ and Er$_{2}$Zr$_{2}$O$_{7}$, and can be attributed to the paramagnet-like behavior down to 1.8 K. A similar behavior is reported for other disordered pyrochlore oxides Dy$_{2}$Zr$_{2}$O$_{7}$, Ho$_{2}$Zr$_{2}$O$_{7}$ and clean pyrochlores Dy$_{2}$Ti$_{2}$O$_{7}$, Ho$_{2}$Ti$_{2}$O$_{7}$, etc. The Eu$_{2}$Zr$_{2}$O$_{7}$ shows an irreversibility in magnetization data below $\sim$90 K, similar to other Eu based compounds Eu$_{2}$Ti$_{2}$O$_{7}$, Eu$_{2}$CoO$_{4}$, etc \cite{dasgupta2007low,simovivc2003local,tovar1989eu}. Below this temperature, the system exhibits a plateau region that lasts down to 20 K. The plateau region or the nearly constant susceptibility between T = 20 - 90 K is observed even for the higher magnetic field of 50 kOe (see Fig. 5) and is known to arise from the Van Vleck paramagnetic contribution of the Eu$^{3+}$ ions \cite{morse1932theory, rettori1996esr}. The free Eu$^{3+}$ ion with an electronic configuration of 4\textit{f}$^{6}$ has a nonmagnetic ground state $^{7}$\textit{F}$_{0}$ which is mixed with the first excited magnetic state $^{7}$\textit{F}$_{1}$ in the presence of the external magnetic field, thereby leading to the considerable contribution of the zeroth Van Vleck term. This term is constant at low temperatures but due to the excited state's thermal population, it exhibits temperature-dependent behavior for T $\geq$ 90 K, by adding the Curie-Weiss contribution to the Van Vleck. The anisotropic behavior of Eu$^{3+}$ is attributed to the crystal field splitting of the $^{7}$F$_{J}$ ground states \cite{hundley1989specific}. 
	
	\begin{figure}[htbp]
		\begin{center}
			\includegraphics[scale=0.55]{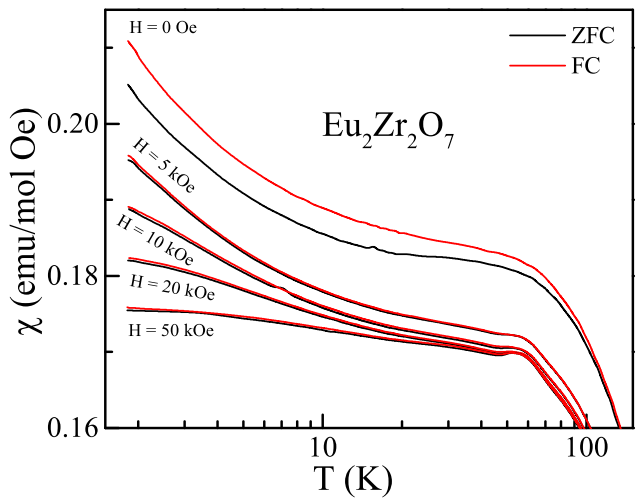}
			\vspace{-5pt}
			\caption{ Dc magnetization (zero field cooled and field cooled) as a function of temperature at various magnetic fields for Eu$_{2}$Zr$_{2}$O$_{7}$.}
		\end{center}
		\vspace{-25pt}
	\end{figure} 
	
	The presence of magnetic frustration can be inferred from the values of Curie-Weiss temperature ($\theta_{CW}$) derived from the Curie-Weiss fit of the inverse susceptibility data above 30 K (Fig. 4 (right). The obtained $\theta_{CW}$ for Gd$_{2}$Zr$_{2}$O$_{7}$ and Er$_{2}$Zr$_{2}$O$_{7}$ are -11.82(3) K and -12.09(02) K which are closer to that of their titanates analogues \cite{luo2001low,vlavskova2019evidence}. The negative sign of $\theta_{CW}$ indicates the dominance of antiferromagnetic correlations and the absence of ordering temperature shows the strength of magnetic frustration present in these systems. The calculated effective magnetic moment ($\mu_{eff}$) from Curie constant are found to be 11.28 (17) $\mu_{B}$/Eu, 5.88 (19) $\mu_{B}$/Gd and 6.58 (02) $\mu_{B}$/Er are significantly lesser than their free ion values (in case of Gd, Er) pointing to the effect of lattice disorder . A significant large value of $\mu_{eff}$ in Eu$_{2}$Zr$_{2}$O$_{7}$ is presumable due to the contribution of Van Vleck paramagnetism \cite{pal2018high}.
	
	Figure 6 shows the isothermal magnetization measured at 2 K for the three compounds showing no sign of phase transition. At the highest applied field of $\textit{H}$ = 70 kOe, the magnetic moment per Gd$^{3+}$ ion is $\mu$ $\approx$ 4.8$\mu_{B}$/Gd, which is $\sim$ 70$\%$ of the theoretically expected saturation moment of $\mu_{sat}$ = 7.0$\mu_{B}$ for spin S = 7/2. The isotherm follows the linear behavior up to the field of approximately 7 kOe until saturation effects set in. This behavior is in contrast to the M(\textit{H}) dependence for free Gd$^{3+}$ ions with a large spin value of S = 7/2, which shows a rapid saturation at such a low temperature, indicating the significance of antiferromagnetic interactions in the system at 2 K. The linear behavior of isotherm for Eu$_{2}$Zr$_{2}$O$_{7}$ 2K suggesting the dominance of antiferromagnetic interactions in the system. The absence of saturation magnetization and the reduced value of maximum achieved magnetic moment at the highest measurement field reveals the presence of strong crystal field anisotropy in these compounds. 
	
	\begin{figure}[htbp]
		\begin{center}
			\includegraphics[scale=0.5]{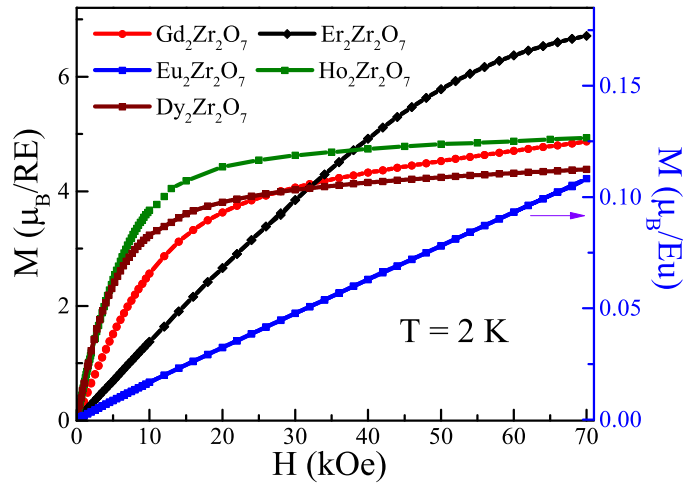}
			\vspace{-5pt}
			\caption{(a) Isothermal magnetization as a function of magnetic field for A$_{2}$Zr$_{2}$O$_{7}$ (A = Er, Gd, Eu, Ho, Dy) at \textit{T} = 2 K. The magnetization data for Ho and Dy zirconate is taken from reference 13 $\&$ 14.}
		\end{center}
		\vspace{-25pt}
	\end{figure} 
	
	\subsection[A]{Ac susceptibility}
	
	\begin{figure}[htbp]
		\begin{center}
			\includegraphics[scale=0.43]{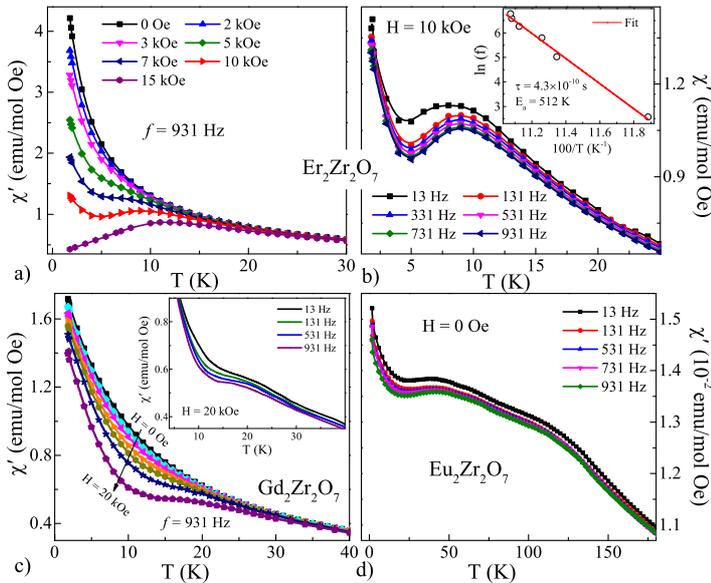}
			\caption{(a) Temperature dependence of real part of ac susceptibility ($\chi^{\prime}$) of Er$_{2}$Zr$_{2}$O$_{7}$ at 931 Hz in various fields 0 $\leq$ \textit{H} $\leq$ 15 kOe. (b) Temperature variation in $\chi^{\prime}$ measured at different frequencies (13 $\leq$ \textit{f} $\leq$ 931 Hz at \textit{H} = 10 kOe. Inset: plot of ln(\textit{f}) versus 1/\textit{T}, where \textit{T} is the temperature estimated from ac susceptibility data, (red) solid line is the best fit using the Arrhenius law. (c) Ac susceptibility data of Gd$_{2}$Zr$_{2}$O$_{7}$ in the field range (0 - 20 kOe). Inset: shows the ac susceptibility data at different frequencies at \textit{H} = 20 kOe. (d) Zero field ac susceptibility data of Eu$_{2}$Zr$_{2}$O$_{7}$ in the frequency range (0 - 1 kOe).}
		\end{center}
	\end{figure} 
	
	Figure 7 shows the ac susceptibility data for all the compounds. The real and imaginary part of ac susceptibility were measured at various frequencies (0 - 1 kOe) and fields (0 - 20 kOe) values. However, due to the large noise in the imaginary part, only the real part is analyzed. Er$_{2}$Zr$_{2}$O$_{7}$ and Gd$_{2}$Zr$_{2}$O$_{7}$ exhibits paramagnetic-like behavior down to 1.8 K at \textit{H} = 0 Oe. On adding the dc magnetic field the susceptibility curve deviate from the paramagnetic regime and shows a weak anomaly around \textit{T$_{\textit{f}}$} $\approx$ 10 K at \textit{H} = 7 kOe and 10 kOe for Er$_{2}$Zr$_{2}$O$_{7}$ and Gd$_{2}$Zr$_{2}$O$_{7}$ respectively. The peak becomes more prominent and shifts towards high temperature with a further increase in the magnetic field. The shift in peak temperature is studied by calculating the Mydosh parameter (\textit{p}) and Arrhenius fit. The Mydosh parameter is calculated using the relation \textit{p} = $\Delta$T$_{\textit{f}}$/T$_{\textit{f}}$$\Delta$(ln$\textit{f}$), where T$_{\textit{f}}$ is the peak temperature corresponding to the measuring frequencies and $\Delta$T$_{\textit{f}}$ is the difference between the peak positions measured at frequencies separated by $\Delta$(ln$\textit{f}$) frequency. The value of \textit{p} obtained for Er$_{2}$Zr$_{2}$O$_{7}$ is 0.017, which is close to the value expected for the typical spin-glass system (\textit{p} = 0.005 - 0.01) \cite{mydosh1993spin}. Further we fit the data with the Arrhenius equation (see inset of Fig. 7b) given by $\tau$ = $\tau_{o}$exp(E$_{a}$/k$_{B}$T$_{\textit{f}})$, where E$_{a}$, k$_{B}$, $\tau$ represents the activation energy, Boltzmann constant and relaxation time respectively. This fit yields parameters value as $\tau$ $\sim$ 10$^{-10}$ s and E$_{a}$ = 512(34) K which is in close agreement to spin glass systems ($\tau$ $\sim$ 10$^{-11}$ - 10$^{-13}$ s) \cite{mydosh1993spin,vincent2018spin}. However, the characterization of the spin-glass state by fit to these models is somewhat inclusive due to the absence of irreversibility in dc magnetization. From these study, it can be inferred that the disordered pyrochlore oxide Er$_{2}$Zr$_{2}$O$_{7}$ exhibits field-induced spin freezing behavior not lying in the spin-glass regime. We are unable to study the field-induced magnetic anomaly of Gd$_{2}$Zr$_{2}$O$_{7}$ due to the lack of clear peak up to the maximum applied field of H = 20 kOe. Unlike Eu$_{2}$Ti$_{2}$O$_{7}$, Eu$_{2}$Zr$_{2}$O$_{7}$ shows an unusual behavior in ac susceptibility without any frequency dependence \cite{pal2018high}. The underlying freezing mechanism of the pyrochlore oxide is associated to the geometrical frustration of rare-earth spin and the anisotropy originated from crystal field of the system, which is not apparently observed in structurally disordered Eu$_{2}$Zr$_{2}$O$_{7}$.
	
	\subsection[A]{Heat Capacity}
	
	\begin{figure}[htbp]
		\begin{center}
			\includegraphics[scale=0.55]{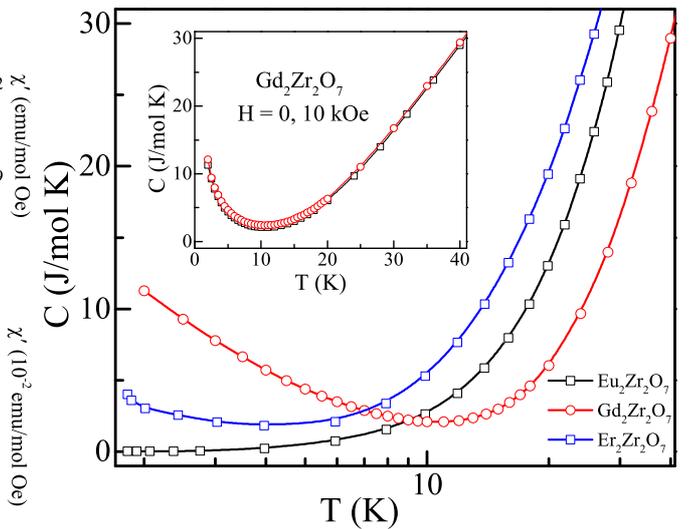}
			\caption{Heat capacity as a function of temperature for 1.8 $\leq$ \textit{T} $\leq$ 40 measured in zero field for A$_{2}$Zr$_{2}$O$_{7}$ (A = Eu, Gd, Er). Inset: Heat capacity data of Gd$_{2}$Zr$_{2}$O$_{7}$ at \textit{H} = 0 (black), 10 kOe (Red).}
		\end{center}
	\end{figure} 
	
	The heat capacity C(\textit{T}) data of A$_{2}$Zr$_{2}$O$_{7}$ (A = Eu, Gd, Er) are shown in Fig. 8. Like the dc susceptibility the C(\textit{T}) data do not show any anomaly related to any transition down to 1.8 K. However, an increase in the C(\textit{T}) is observed in A$_{2}$Zr$_{2}$O$_{7}$ (A = Gd, Er) with decreasing temperature below 10 K might be the precursor of magnetic ordering. This behavior is similar to that of Dy$_{2}$Zr$_{2}$O$_{7}$, Dy$_{1.7}$La$_{0.3}$Zr$_{2}$O$_{7}$ (disordered pyrochlores) and Dy$_{2}$Ti$_{2}$O$_{7}$, Dy$_{1.8}$Ca$_{0.2}$Ti$_{2}$O$_{7}$, Nd$_{2}$Zr$_{2}$O$_{7}$ (perfect pyrochlores), indicating the development of short-range correlations at low temperature \cite{sheetal2020emergence,sheetal2021evolution,anand2015investigations,lutique2003low}. Inset of Fig. 8 shows the C(\textit{T}) data of Gd$_{2}$Zr$_{2}$O$_{7}$ at \textit{H} = 0, 10 kOe. Despite the spin freezing anomaly in ac susceptibility, no appreciable change in the position of upturn and magnitude of C(\textit{T}) is observed. This rules out the possibility of spin-glass transition in both the samples. The absence of spin ice state, with a large entropy of $\sim$ Rln2 was reported for the disordered pyrochlore oxide Dy$_{2}$Zr$_{2}$O$_{7}$ \cite{ramon2019absence}. Our heat capacity measurements are limited to 1.8 K, but with a similar structural disorder, a dynamic ground state with large entropy would be expected for these compounds. A recent study on Er$_{2}$Zr$_{2}$O$_{2}$ shows a very dynamic ground state with an entropy of Rln2 attributed to the modified crystal field state due to the structural symmetry of disordered pyrochlore \cite{vlavskova2019evidence}. Unlike Er$_{2}$Zr$_{2}$O$_{2}$ and Gd$_{2}$Zr$_{2}$O$_{2}$ compounds, Eu$_{2}$Zr$_{2}$O$_{2}$ does not show any upturn in C(\textit{T}) at low temperature down to 1.8 K as expected for the $^{7}$F$_{0}$ electronic ground state of the system as evident in dc magnetic susceptibility data as well. 	
	
	\section{Discussion}

	The A$_{2}$Zr$_{2}$O$_{7}$ (A = Er, Gd, Eu) systems crystallize in disordered pyrochlore structure in accordance with their r$_{A}$/r$_{B}$ radius ratio, like other Dy/Ho based zirconate oxides. The increased A$^{3+}$/B$^{3+}$ disorder in the pyrochlore structure impact the magnetic ground-state of these systems. In diluted systems, the magnetic or nonmagnetic dopant randomly distributed over the lattice acts as a host for the various exotic states \cite{anand2015investigations,ke2007nonmonotonic,gomez2014spin}. The pyrochlore-like structural ordering of the disordered zirconates leads to  the significant geometrical frustration effects at the local level, which are further enhanced by the presence of random-bond disorder caused by the dilution \cite{gomez2014spin}. From the magnetization studies we found that A$_{2}$Zr$_{2}$O$_{7}$ (A = Er, Gd) exhibit paramagnet-like behavior and the reduced values of $\mu_{eff}$ and saturation magnetization indicate the presence of strong crystal field anisotropy in these systems. Compared to Dy, Ho, Eu based pyrochlore zirconate, the crystal field effect are more dominant in Er and Gd based pyrochlore zirconate. This is consistent with the structural distortion in cubic geometry introduced by the distribution of oxygen around Rare earth site.\\  
	
	In Er$_{2}$Zr$_{2}$O$_{7}$, a spin glass state is reported below 1 K \cite{vlavskova2019evidence} and the system is remain dynamic down to 0.8 K with an entropy of Rln2 expected for the maximum possible 2$^{N}$ state. However, on applying an external magnetic field an unusual spin freezing at high temperature is observed in Eu$_{2}$Zr$_{2}$O$_{7}$ and Gd$_{2}$Zr$_{2}$O$_{7}$. The disordered pyrochlore A$_{2}$Zr$_{2}$O$_{7}$ (A = Dy, Ho) also show similar behavior in ac susceptibility at high temperature. The emergence of field-induced spin freezing on applying external magnetic field is presumably due to the slowing down of spin dynamics is these structurally disordered systems. Ehlers \textit{et al.} discussed the possible origin for the observation of high temperature spin freezing in the spin ice Dy$_{2}$Ti$_{2}$O$_{7}$ in addition to the spin-glass-like state below 1 K \cite{ehlers2002dynamical} in both the compounds.     
	
	Er$_{2}$Zr$_{2}$O$_{7}$ and Gd$_{2}$Zr$_{2}$O$_{7}$ shows an upturn in the heat capacity data below 20 K, which may eventually form a full peak at low-temperature below 1.8 K. Although we could not observe full peak due to our limited temperature range accessibility, we expect the low temperature heat capacity behavior to be similar to other nominal ordered/disordered pyrochlores Dy$_{2}$Ti$_{2}$O$_{7}$, Er$_{2}$Ge$_{2}$O$_{7}$, (Ce/Dy)$_{2}$Zr$_{2}$O$_{7}$, that show similar broad feature in heat capacity below 1 K associated to the unconventional spin dynamics \cite{snyder2001spin,dun2015antiferromagnetic,gao2019experimental, sheetal2020emergence}. The Schottky contribution to the low temperature heat capacity an energy gap of 0.17 meV to the first excited state in Er$_{2}$Zr$_{2}$O$_{7}$ \cite{ehlers2002dynamical}. Whereas, the inelastic neutron scattering studies revealed that the low temperature anomaly is not of Schottky type and the structural difference between the pyrochlore and disordered pyrochlore modifies the effect of crystal field scheme in this compound \cite{gaudet2018effect}. Thus the upturn trend in heat capacity of Er$_{2}$Zr$_{2}$O$_{7}$ and Gd$_{2}$Zr$_{2}$O$_{7}$ might have the same origin as in the case of Dy$_{2}$Zr$_{2}$O$_{7}$ \cite{sheetal2020emergence}.
	
	\section{Conclusion}

	In conclusion, like other zirconates Dy$_{2}$Zr$_{2}$O$_{7}$, Ho$_{2}$Zr$_{2}$O$_{7}$, the A$_{2}$Zr$_{2}$O$_{7}$ (A = Er, Gd, Eu) were confirmed to crystallized in disordered pyrochlore structure. The large susceptibility value and the absence of spin freezing in zero field is consistent with the structural behavior of these systems. A$_{2}$Zr$_{2}$O$_{7}$ (A = Er, Gd) shows the field induced spin freezing behavior whereas Eu$_{2}$Zr$_{2}$O$_{7}$ do not show spin freezing owing to its non-magnetic ground state. The lack of irreversibility in dc magnetization and the heat capacity studies rules out the glassy behavior in Er$_{2}$Zr$_{2}$O$_{7}$. Our measurements indicate the absence of long-range ordering in these systems. On comparing with the existing literature of the same family of compounds, a dynamic ground state is expected for these systems, however the advanced local probe measurements would be useful to establish the true ground state of these systems.\\
	
	\textbf{Acknowledgement:} We thank AMRC, Indian Institute of Technology Mandi for the experimental facility. Sheetal Acknowledged IIT Mandi and MHRD India for the HTRA fellowship.

	\bibliography{Ref}
\end{document}